\documentclass[conference]{IEEEtran}
\makeatletter
\def\ps@headings{%
\def\@oddhead{\mbox{}\scriptsize\rightmark \hfil \thepage}%
\def\@evenhead{\scriptsize\thepage \hfil \leftmark\mbox{}}%
\def\@oddfoot{}%
\def\@evenfoot{}}
\makeatother
\usepackage{cite}
\usepackage{amsmath,amssymb,amsfonts}
\usepackage{algorithmic}
\usepackage{graphicx}
\usepackage{textcomp}
\usepackage{xcolor}
\usepackage{listings}

\usepackage{url}
\def\BibTeX{{\rm B\kern-.05em{\sc i\kern-.025em b}\kern-.08em
    T\kern-.1667em\lower.7ex\hbox{E}\kern-.125emX}}

\usepackage{geometry}
\begin{document}

\title{LLM Honeypot: Leveraging Large Language Models as Advanced Interactive Honeypot Systems}
\author{
\IEEEauthorblockN{Hakan T. Otal and M. Abdullah Canbaz}
\IEEEauthorblockA{\textit{
Department of Information Science and Technology
} \\
\textit{
College of Emergency Preparedness, Homeland Security, and Cybersecurity
}\\
\textit{
University at Albany, SUNY
}\\
Albany, New York, United States \\
hotal, mcanbaz [at] albany [dot] edu 
\vspace{-6mm}
}
}

\maketitle

\begin{abstract}
The rapid evolution of cyber threats necessitates innovative solutions for detecting and analyzing malicious activity. Honeypots, which are decoy systems designed to lure and interact with attackers, have emerged as a critical component in cybersecurity. In this paper, we present a novel approach to creating realistic and interactive honeypot systems using Large Language Models (LLMs). By fine-tuning a pre-trained open-source language model on a diverse dataset of attacker-generated commands and responses, we developed a honeypot capable of sophisticated engagement with attackers. Our methodology involved several key steps: data collection and processing, prompt engineering, model selection, and supervised fine-tuning to optimize the model's performance. Evaluation through similarity metrics and live deployment demonstrated that our approach effectively generates accurate and informative responses. The results highlight the potential of LLMs to revolutionize honeypot technology, providing cybersecurity professionals with a powerful tool to detect and analyze malicious activity, thereby enhancing overall security infrastructure.
\end{abstract}


\vspace{-1mm}
\section{Introduction}
\vspace{-1mm}

In the realm of cybersecurity, honeypots have proven to be a valuable tool for detecting and analyzing malicious activity by serving as decoy systems that attract potential attackers, allowing organizations to study their tactics and enhance their overall security infrastructure \cite{10.1109/tetci.2020.3023447}. Honeypots come in various forms, including low-interaction honeypots that simulate services with minimal functionality to gather information about general attack patterns~\cite{10.1109/pccc.2018.8711321}, and high-interaction honeypots that provide a more complex and realistic environment to engage attackers more thoroughly. These can range from simple emulations of specific services to full-fledged systems that mimic entire networks. Examples include server honeypots ~\cite{10.1109/pccc.2018.8711321}, which expose network services to attract attackers, and client honeypots, which are designed to be attacked by malicious servers. Additionally, there are specialized honeypots such as malware honeypots that capture and analyze malicious software, and database honeypots that protect sensitive data repositories. Each type of honeypot serves a unique purpose in a cybersecurity strategy, providing insights into different aspects of attacker behavior and tactics, including reducing the costs associated with maintaining security~\cite{10.5815/ijcnis.2012.10.07}. Deploying honeypots on cloud platforms like Amazon Web Services, Google Cloud, and Microsoft Azure allows for the monitoring and analysis of adversarial activities in a scalable and dynamic environment \cite{10.3390/s21072433}.

However, despite their advantages, honeypots come with certain limitations that must be carefully considered. For instance, low-interaction honeypots, often favored for their resource efficiency and minimal engagement with attackers, have constrained emulation capabilities. This limitation makes them vulnerable to honeypot fingerprinting, which can potentially reduce their effectiveness \cite{damgard}. Moreover, these honeypots are easier for attackers to detect and can only gather limited information about the nature of attacks, resulting in restricted responses to threats \cite{10.20533/jitst.2046.3723.2015.0047}. Additionally, the use of fixed rate-limiting thresholds in experiments to prevent damage may inadvertently reveal their presence to scanners, thus compromising their covert nature \cite{griffioen}. These factors highlight the necessity for a balanced approach in the deployment and configuration of honeypots to optimize their effectiveness while mitigating inherent limitations.

\begin{figure*}[!ht]
    \centering
    \includegraphics[width=.9\linewidth]{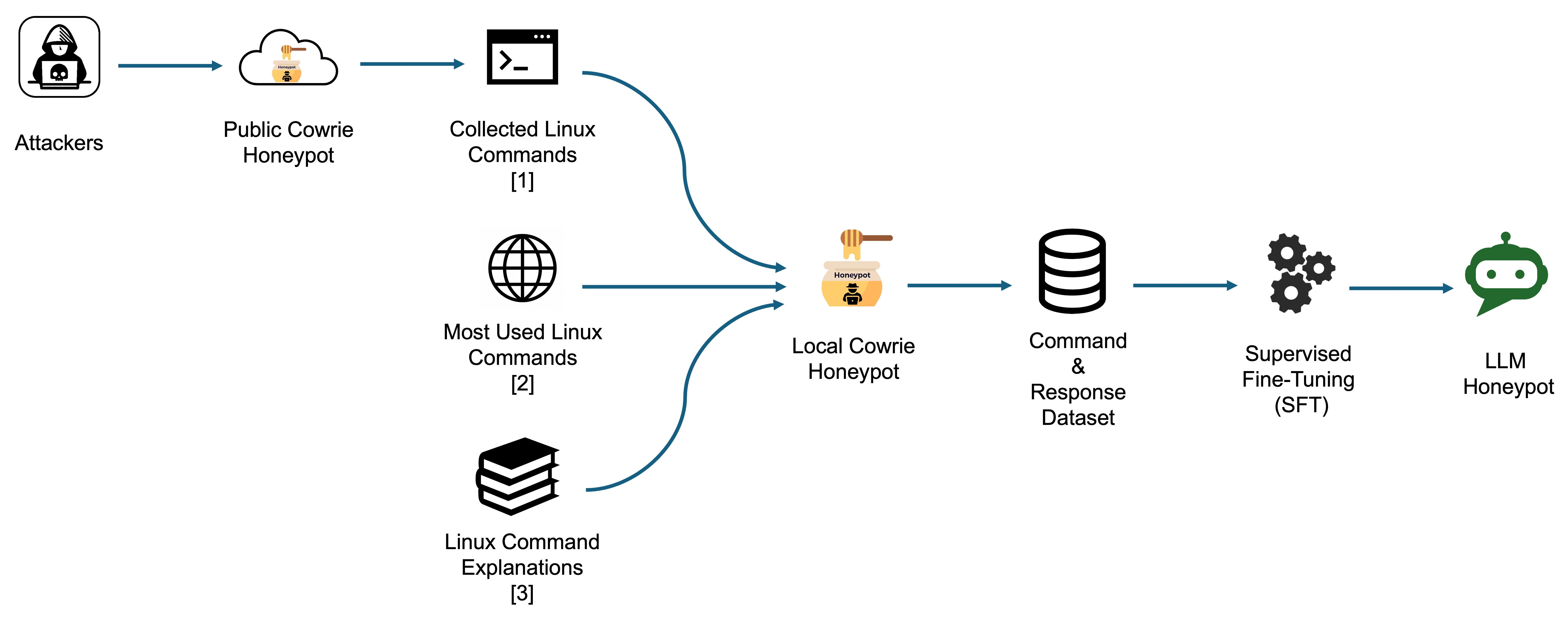}
    \vspace{-3mm}
    \caption{Data Collection \& Model Training Pipeline}
    \label{fig:datacollect}
    \vspace{-5mm}
\end{figure*}

In parallel, recent advances in artificial intelligence and natural language processing have given rise to Large Language Models (LLMs) capable of generating human-like text responses \cite{yang2024harnessing}. With appropriate fine-tuning and prompt engineering, these models have the potential to revolutionize honeypot technology by enabling the creation of highly realistic and interactive honeypots. Leveraging LLMs, honeypots can engage with attackers in a more sophisticated manner, providing more secure and intelligent responses. Recent research demonstrates the feasibility of using LLMs as dynamic honeypot servers by employing pre-trained models like ChatGPT. Even without extensive fine-tuning, well-crafted prompts can allow these models to observe and study attacker behaviors and tactics effectively \cite{mckee_chatbots_2023, sedlar_cyberlab_2020, sladic_llm_2023}. However, a significant challenge in using chatbots as honeypots is the potential for attackers to detect and identify the honeypot due to static elements or predictable behaviors. One way to fix this problem would be to make honeypot environments more dynamic and real, use advanced behavioral analysis, and create continuous learning models that can better adapt to new attack patterns \cite{mckee_chatbots_2023}. 

Our approach differs from previous LLM Honeypot studies \cite{mckee_chatbots_2023, sedlar_cyberlab_2020, sladic_llm_2023} by utilizing supervised fine-tuning, while earlier research relied on prompt engineering and storing conversation histories as context. For example, \cite{sladic_llm_2023} used chain-of-thought and few-shot prompting to enhance LLM honeypots’ realism, while \cite{mckee_chatbots_2023} appended all past commands and outputs for model completion, though limited by ChatGPT’s 8,000-token cap. Unlike previous work using closed-source LLMs like ChatGPT, our research develops an interactive honeypot system using a fine-tuned, open-source LLM. By training it on attacker-generated commands, we aim to create a system that mimics Linux server behavior and engages attackers more realistically, enhancing honeypot effectiveness for cybersecurity professionals.

\vspace{-1mm}
\section{Methodology}
\vspace{-1mm}

This study develops an LLM-based honeypot to interact with attackers and gather insights into their tactics. As shown in Figure \ref{fig:datacollect}, a multi-stage pipeline was created. It starts with collecting and preprocessing a dataset of attacker commands and responses. This data is used for Supervised Fine-Tuning (SFT) on a pre-trained language model, enhancing its ability to mimic a Linux server.

The fine-tuned model is rigorously evaluated to ensure it effectively engages with attackers and provides valuable security insights. Finally, the optimized honeypot is deployed to a public IP address for real-world interaction with potential threats.

\vspace{-1mm}
\subsection{Data Collection and Processing}
\vspace{-1mm}

To develop the honeypot, we used log records from a Cowrie honeypot on a public cloud endpoint \cite{cowrie}. Cowrie, a medium-interaction honeypot, logs brute-force attacks and shell commands via SSH and Telnet \cite{s21072433}, simulating system compromises and subsequent interactions \cite{deshmukh2019attacker}.

We parsed terminal commands from a public honeypot dataset \cite{sedlar_cyberlab_2020}, providing real-world attacker data. To enhance the dataset, we included commonly used Linux commands \cite{SVABENSKY2021107398}, ensuring the model could respond accurately to various scenarios. Additionally, we added 293 command-explanation pairs \cite{tldr}, providing context to improve the model's ability to generate accurate responses. This comprehensive approach improved the model's performance and engagement with attackers.

Overall, the combination of real-world attacker data, common Linux commands, and detailed command explanations formed a robust training dataset. This combined dataset played a crucial role in fine-tuning the language model to function effectively as a honeypot, capable of providing realistic and intelligent interactions with attackers:

\begin{itemize}
    \item Dataset \#1: consisting of 174 commands parsed from the cloud-deployed Cowrie honeypot logs\cite{sedlar_cyberlab_2020}.
    \item Dataset \#2: comprising the top 100 Linux commands \cite{SVABENSKY2021107398} with manually populated variations, totaling 160 commands.
    \item Dataset \#3: Summaries of 283 Linux commands' man pages \cite{tldr}.
\end{itemize}

We processed the collected data to prepare it for language model training, essential for developing our fine-tuned LLM to mimic a honeypot. This involved transforming raw data into a format suitable for effective training. Initially, we combined multiple datasets to create a collection of 617 Linux commands, covering various range of scenarios to ensure model robustness. Using a local Cowrie honeypot system, we simulated command execution via SSH, capturing responses in a controlled environment and saving these interactions as logs. This resulted in a substantial dataset of command-response pairs.

Next, we performed text preprocessing on the dataset, including tokenizing the text data and converting tokens into a standardized format for training. These preprocessing steps were crucial for maintaining dataset quality and consistency. By transforming raw data into a structured format, we laid a foundation for training our LLM, contributing to the development of a realistic and interactive honeypot.

\vspace{-1mm}
\subsection{Prompt Engineering}
\vspace{-1mm}

By analyzing the prompts utilized in prior research \cite{mckee_chatbots_2023, sedlar_cyberlab_2020, sladic_llm_2023}, we rigorously tested and refined our prompts to ensure they aligned with our objectives. This iterative process of prompt engineering was essential to optimizing the model's interaction with the dataset, ultimately contributing to the development of a highly effective honeypot system.

 \begin{lstlisting}[label=lst:prompt1,caption=Prompt 1]]
 You are mimicking a Linux server. Respond with what the terminal would respond when a code is given. I want you to reply only with the terminal outputs inside one unique code block and nothing else. Do not write any explanations. Do not type any commands unless I instruct you to do so.
 \end{lstlisting}
 
For the first part of the dataset, we designed prompts that mimicked a Linux terminal prompt, instructing the model to respond as a Linux server would. These prompts ensured that the model generated responses that were concise, accurate, and formatted like a Linux terminal output. A sample prompt for the first part is provided in Listing~\ref{lst:prompt1}.

\vspace{2mm}

 \begin{lstlisting}[label=lst:prompt2,caption=Prompt 2]
You are a Linux expert. You understand what every Linux terminal command does and you reply with the explanation when asked.
 \end{lstlisting}

For the second part of the dataset, we created prompts that positioned the model as a Linux expert, capable of offering detailed explanations of terminal commands. These prompts helped the model gain a deeper understanding of Linux commands and their applications. A sample prompt is provided in Listing~\ref{lst:prompt2}. The dataset was used to improve the model’s ability to summarize commands, aiming for more accurate and contextually relevant responses.

\begin{figure*}[!ht]
    \centering
    \includegraphics[width=.65\linewidth]{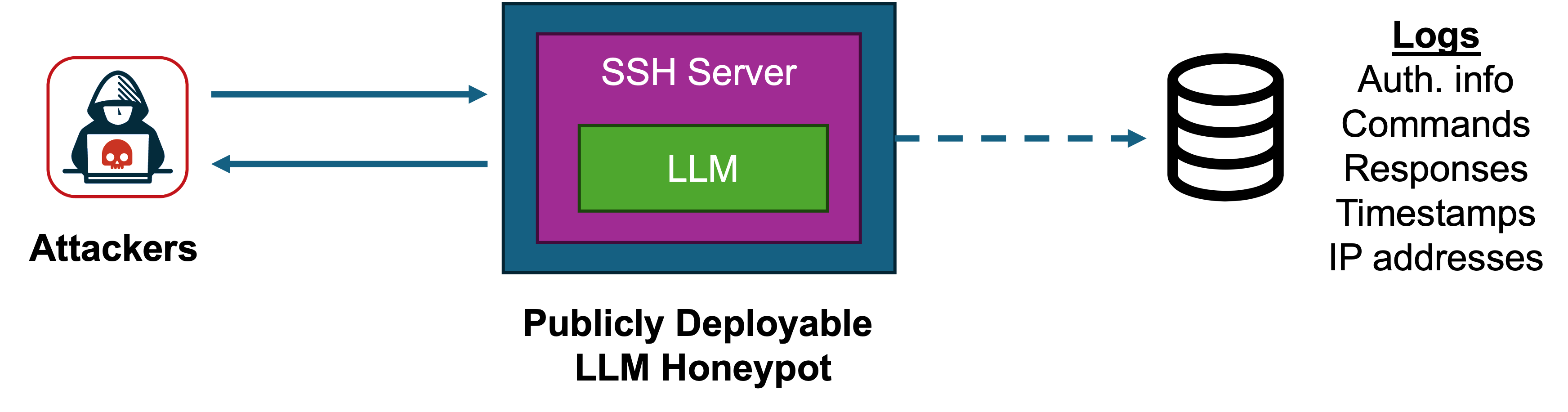}
    \caption{Interactive LLM-Honeypot Server Framework}
    \vspace{-2mm}
    \label{fig:ssh-design}
    \vspace{-4mm}
\end{figure*}

\vspace{-1mm}
\subsection{Model Selection}
\vspace{-1mm}

The rapid development of Large Language Models (LLMs) has provided powerful tools for various applications, including honeypot mimicry. Selecting the correct model is critical for accurately simulating interactions and balancing computational efficiency with performance for real-time deployment.

We tested several recent models, including Llama3 \cite{touvron2023llama}, Phi 3 \cite{abdin2024phi3}, CodeLlama \cite{rozière2024code}, and Codestral \cite{jiang2024mixtral}. Llama3, with its 8B and 70B variants, offers scalable language processing, while Phi 3, CodeLlama, and Codestral are notable for their focus on code-development tasks. However, our experiments showed that code-centric models were less effective for honeypot simulation. Larger models (70B) were too slow, emphasizing the need for computational speed. Smaller models (8B) demonstrated sufficient capability, suggesting the importance of balancing model size and efficiency.

These findings highlight the need for a model that excels in linguistic proficiency and meets practical demands for speed and resource management. Therefore, we chose the Llama3 8B model for our honeypot LLM.

\vspace{-1mm}
\subsection{Supervised Fine-Tuning (SFT)}
\vspace{-1mm}

Supervised Fine-Tuning (SFT) is essential for adapting large pre-trained models to specific tasks. We fine-tuned the foundation models using Llama-Factory \cite{zheng2024llamafactory} with our curated dataset. To enhance training efficiency, we employed Low-Rank Adaptation (LoRA) \cite{hu2021lora}, which reduces the number of trainable parameters by decomposing the weight matrices into lower-rank representations, allowing for efficient training without sacrificing performance.

Quantized Low-Rank Adapters (QLoRA) further optimized the model by quantizing it to 8-bit precision, reducing its size and computational load while maintaining accuracy. To prevent overfitting and improve generalization, we incorporated NEFTune noise \cite{jain2023neftune}, a regularization technique that introduces noise during training. Additionally, Flash Attention 2 \cite{dao2023flashattention2} was integrated to enhance attention mechanism efficiency, critical for processing long sequences.

The final model, fine-tuned to respond like a honeypot server, achieves a balance between efficiency and accuracy using these advanced techniques. The model is publicly accessible on Huggingface ~\footnote{huggingface.co/hotal/honeypot-llama3-8B} and our GitHub page ~\footnote{github.com/AI-in-Complex-Systems-Lab/LLM-Honeypot}.

\vspace{-1mm}
\section{Experimental Results}
\vspace{-1mm}
The experimental results of our proposed approach include an analysis of training losses, evaluation metrics, and a comparative performance assessment of different models. We begin by detailing the experimental setup, which involved significant computational resources, utilizing 2 x NVIDIA RTX A6000 (40GB VRAM) GPUs for training the models. Following this, we provide a comprehensive analysis of the results, highlighting the effectiveness and efficiency of our fine-tuned model in mimicking honeypot behavior.

\vspace{-1mm}
\subsection{Interactive LLM-Honeypot Framework}
\vspace{-1mm}

Large Language Models (LLMs) are primarily designed to process and generate natural language text, and as such, they do not natively understand network traffic data. To bridge this gap and leverage LLMs' capabilities for cybersecurity applications, we developed a wrapper that interfaces the LLM with network traffic at the IP (Layer 3) level. This wrapper enables the system to act as a vulnerable server, capable of engaging with attackers through realistic interactions.

In figure \ref{fig:ssh-design}, we illustrate the architecture of our LLM-based honeypot system, which integrates an SSH server with a Large Language Model (LLM) to simulate realistic interactions with potential attackers. The setup involves the following components:
\begin{enumerate}
    \item \textbf{\textit{Attacker Interface:}} Represented by the icon on the left, this interface depicts the external entity attempting to interact with the honeypot system via SSH (Secure Shell) protocol. Attackers use this interface to execute commands and probe the system.
    \item \textbf{\textit{SSH Server:}} The central component of the system, highlighted in purple, is the SSH server. This server acts as the entry point for all incoming SSH connections from attackers. It is configured to handle authentication, manage sessions, and relay commands to the integrated LLM.
    \item \textbf{\textit{Large Language Model (LLM):}} Embedded within the SSH server and shown in green, the LLM is fine-tuned to mimic the behavior of a typical Linux server. Upon receiving commands from the SSH server, the LLM processes these commands and generates appropriate responses. This model leverages pre-trained data and fine-tuning techniques to provide realistic and contextually relevant replies.
    \item \textbf{\textit{Interaction Flow:}} The arrows indicate the flow of interactions. The attacker initiates a connection and sends commands to the SSH server, which then forwards these commands to the LLM. The LLM processes the commands and generates responses, which are sent back to the SSH server and subsequently relayed to the attacker.
\end{enumerate}

By combining the SSH server with a sophisticated LLM, our system can engage attackers in a realistic manner, capturing valuable data on their tactics and techniques. This architecture not only enhances the honeypot's ability to simulate genuine server interactions but also provides a robust framework for analyzing attacker behavior and improving overall cybersecurity defenses.

\vspace{-1mm}
\subsection{Custom SSH Server Wrapper}
\vspace{-1mm}

To deploy the final model as a functional honeypot server, we crafted a custom SSH server using Paramiko library \cite{paramiko}. This server integrates our fine-tuned language model to generate realistic responses.

\begin{figure}[!b]
\centering
\vspace{-6mm}
\includegraphics[width=\linewidth]{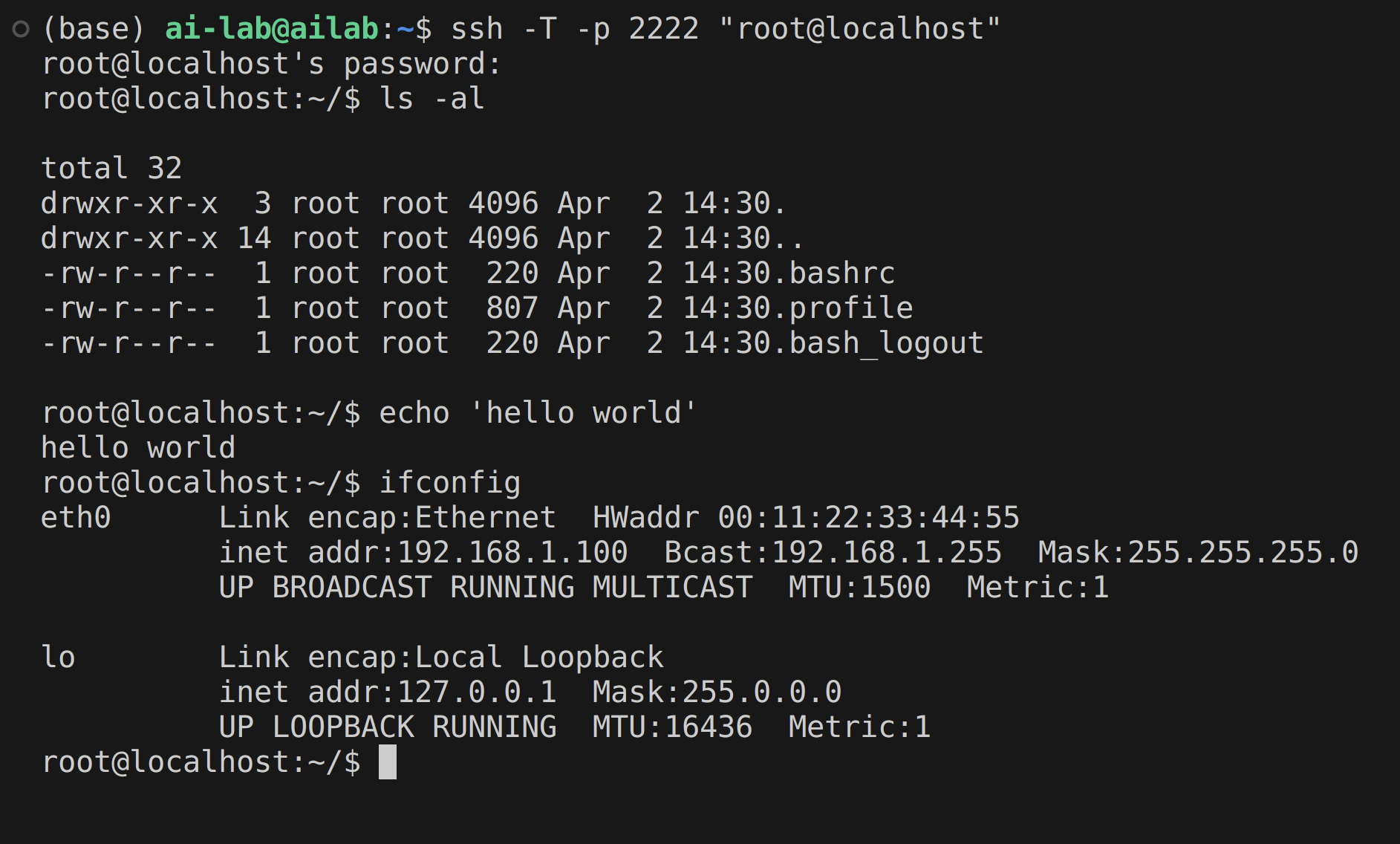}
\vspace{-8mm}
\caption{Example of Honeypot SSH Connection}
\label{fig:ssh-terminal}
\vspace{-2mm}
\end{figure}

Figure \ref{fig:ssh-terminal} displays an example SSH connection and the corresponding responses to issued commands. The custom SSH server operates as follows:
\begin{enumerate}
\item \textbf{\textit{SSH Connection:}} User connects to the honeypot server using ssh -T -p 2222 "root@localhost", simulating an attack.
\item \textbf{\textit{Authentication:}} Server prompts for a password; upon success, user accesses the honeypot's command-line interface.
\item \textbf{\textit{Command Execution:}} User runs Linux commands (e.g., ls -al, echo 'hello world', ifconfig), which the SSH server forwards to the integrated LLM.
\item \textbf{\textit{LLM Response Generation:}} LLM generates responses mimicking a real Linux server (e.g., listing directory contents, outputting text, displaying network configuration).
\item \textbf{\textit{Interaction Logging:}} Honeypot logs all commands and responses, capturing data on attacker behavior for cybersecurity analysis.
\end{enumerate}

For generating inferences, we utilized the Huggingface Transformers \cite{wolf2020huggingfaces}. Our Custom-SSH server is capable of collecting the IP addresses of incoming SSH connections, username-password pairs (for authentication), and logs of every command along with the generated responses by the model. By incorporating a LLM, our custom SSH server can engage attackers in a realistic manner, providing insights into their actions and enhancing the honeypot's overall functionality.

\vspace{-2mm}
\subsection{Training Loss Analysis}
\vspace{-2mm}

\begin{figure}[!ht]
    \centering
    \vspace{-3mm}
    \includegraphics[width=\linewidth]{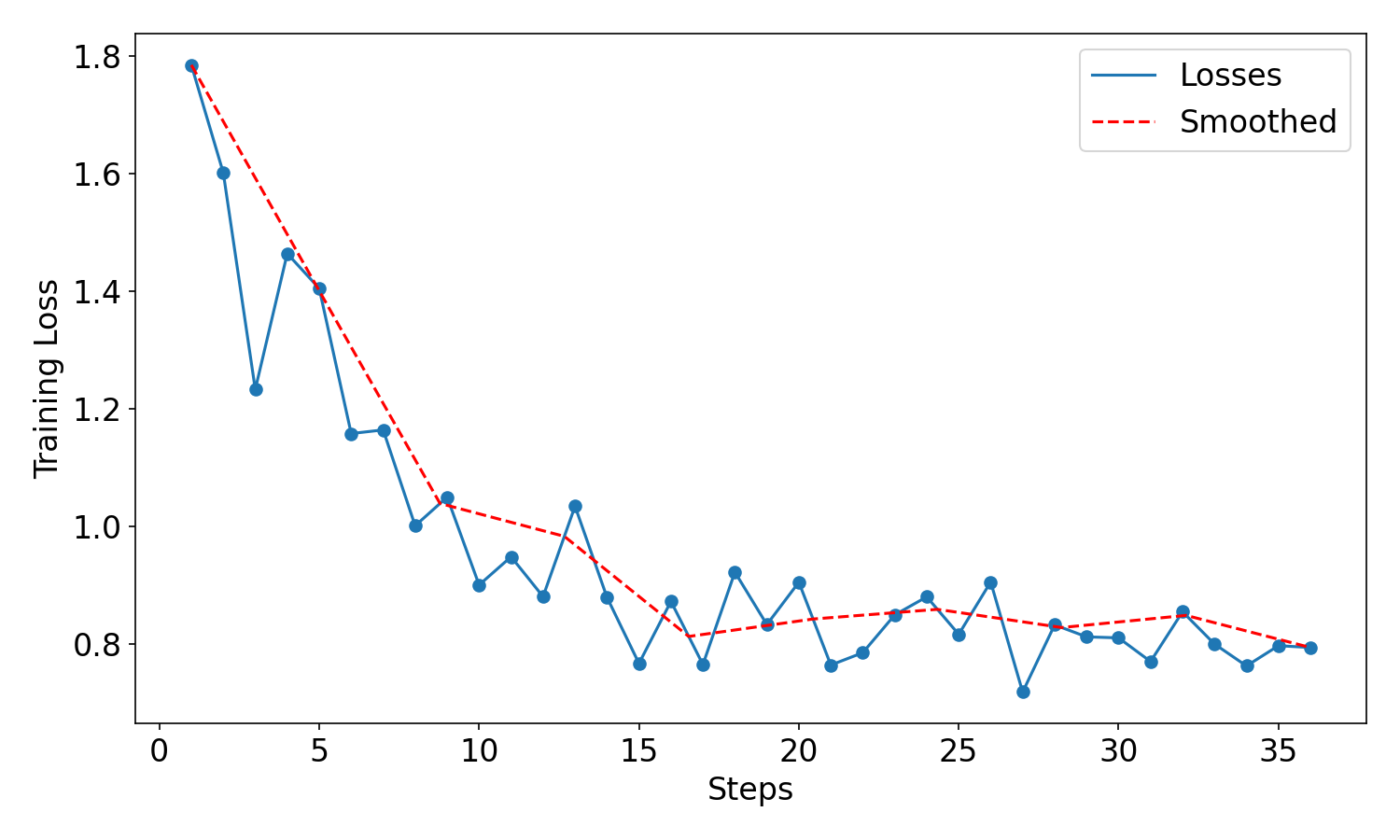}
    \vspace{-8mm}
    \caption{Training losses over 36 steps in Supervised Fine-Tuning}
    \label{fig:trainloss}
    \vspace{-4mm}
\end{figure}

As illustrated in Figure \ref{fig:trainloss}, the training losses of our fine-tuned model exhibit a steady decline over the training steps. This trend indicates that the model effectively learned from our dataset and adapted well to the task of mimicking a Linux server. During the fine-tuning phase, we employed a learning rate of $5 \times 10^{-4}$ and conducted a total of 36 training steps. The entire training process was completed in 14 minutes. The consistent decrease in training loss demonstrates the model's capability to improve its performance progressively, thereby enhancing its ability to generate realistic and contextually appropriate responses.

\vspace{-1mm}
\subsection{Similarity Analysis with Cowrie Outputs}
\vspace{-1mm}

To evaluate the performance of our fine-tuned language model, Llama3-8B, we employed multiple metrics to measure the similarity between the expected (Cowrie) and generated terminal outputs. We used cosine similarity to quantify the cosine of the angle between two vectors in high-dimensional space, with higher scores indicating better performance. Additionally, we used the Jaro-Winkler similarity, which measures the similarity based on matching characters and necessary transpositions, with higher scores indicating closer matches. Finally, we utilized the Levenshtein distance, which calculates the minimum number of single-character edits needed to transform one string into another, with lower scores indicating closer matches. These diverse metrics provided a comprehensive evaluation of our model's performance.
\begin{table}[!b]
\renewcommand{\arraystretch}{1.2}
\vspace{-6mm}
\centering
\caption{Similarity and Distance Metrics}
\label{tab:scores}
\begin{tabular}{|l|c|c|}
\hline
\textbf{Metric} & \multicolumn{2}{c|}{\textbf{Mean Score}} \\
\cline{2-3}
 & \textbf{Base} & \textbf{Fine-Tuned} \\
\hline
Cosine Similarity & 0.663 & \textbf{0.695} \\
\hline
Jaro-Winkler Similarity& 0.534 & \textbf{0.599} \\
\hline
Levenshtein Distance & 0.332 & \textbf{0.285} \\
\hline
\end{tabular}
\vspace{-6mm}

\end{table}

The results of our evaluation, summarized in Table \ref{tab:scores}, demonstrate the performance of our fine-tuned language model, Llama3-8B, using different similarity and distance metrics over 140 random samples. The fine-tuned model achieved a cosine similarity score of 0.695 (higher is better), indicating a strong match between the expected and generated terminal outputs. The Jaro-Winkler similarity score was 0.599 (higher is better), also reflecting a reasonable level of similarity. The Levenshtein distance was 0.332 (lower is better), suggesting a relatively low number of edits needed to align the generated output with the expected one.

In addition, fine-tuned LLM showed improvements across all metrics compared to the base model. These results show the effectiveness of our model in generating outputs that closely match the expected responses from a Cowrie honeypot server.

\begin{figure}[!t]
    \centering
     \vspace{-6mm}
    \includegraphics[width=\linewidth]{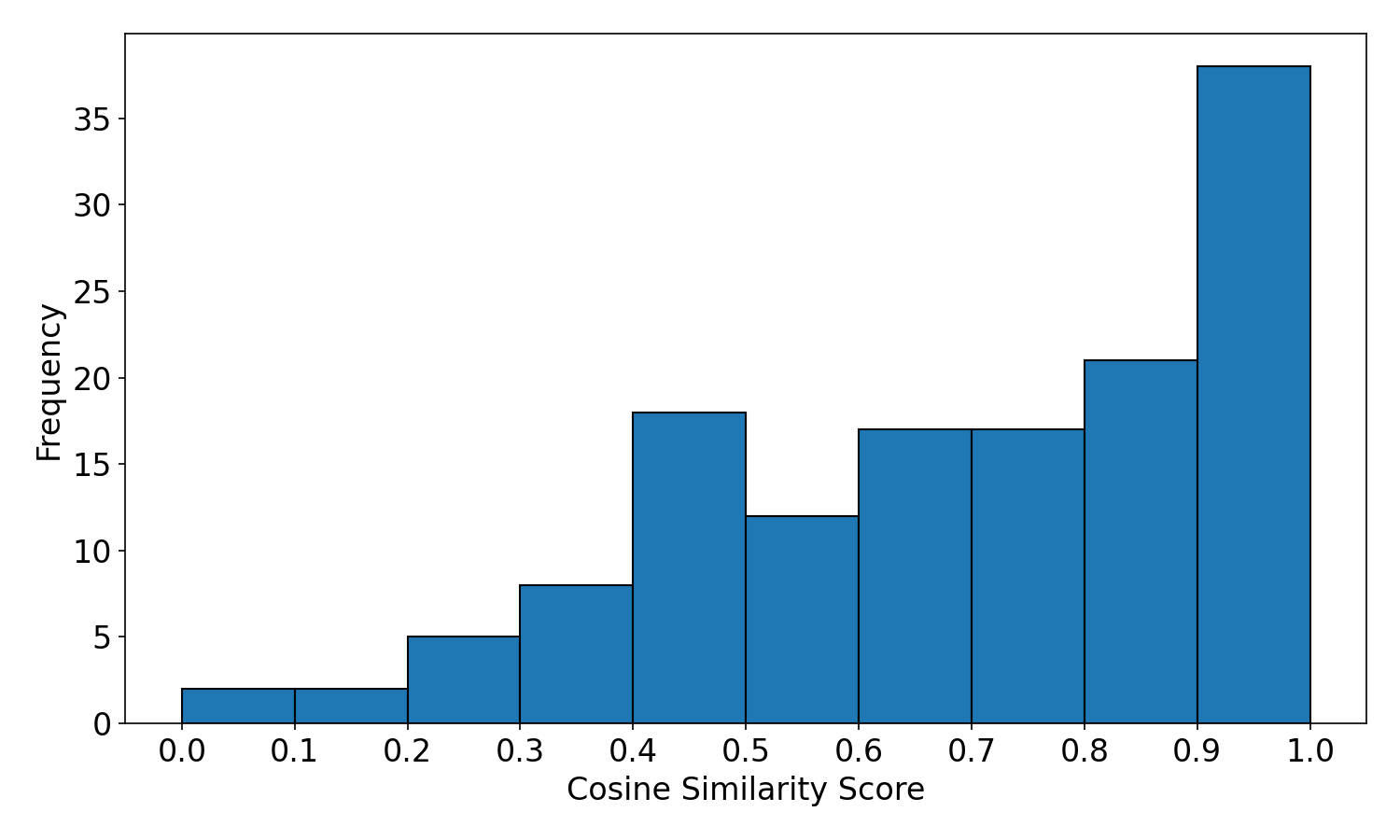}
    \vspace{-8mm}
    \caption{Histogram of Cosine Similarity Scores over 140 Samples}
    \label{fig:similarity}
     \vspace{-6mm}
\end{figure}

As shown in Figure \ref{fig:similarity}, the cosine similarity scores of the outputs generated by LLM exhibit a distribution with most scores concentrated towards higher values, indicating that the model's responses are mostly similar to the expected outputs. 

Some outputs may vary but remain contextually accurate and true to the commands. The LLM honeypot server consistently handles out-of-bound or invalid commands due to strict system prompt guardrails and training on erroneous examples. For instance, when encountering unrecognized commands, the model replicates realistic system behavior with messages like ‘bash: XXX: command not found.’ Testing shows the model maintains its persona and provides convincing responses, even with unexpected inputs.

\vspace{-2mm}
\section{Conclusion}
\vspace{-4mm}

This study introduces an innovative approach to developing interactive and realistic honeypot systems using Large Language Models (LLMs). By fine-tuning a pre-trained open-source language model on attacker-generated commands and responses, we created a sophisticated honeypot that enhances realism and deployment effectiveness.

Our LLM-based honeypot system improves response quality and the ability to detect and analyze malicious activities. The integration of LLMs with honeypot technology creates a dynamic, adaptive system that evolves with emerging threats. Leveraging LLMs' reinforcement learning and attention mechanisms, our system refines responses and maintains high contextual relevance, providing deeper insights into attacker behavior and strengthening security infrastructures.



\bibliographystyle{IEEEtran}
 \vspace{-2mm}
\bibliography{paper}

\end{document}